\documentclass[reprint,amsmath,amssymb,aps,prl,longbibliography,footinbib]{revtex4-1}

\usepackage{graphicx}
\usepackage{dcolumn}
\usepackage{bm}
\usepackage[colorlinks=true, allcolors=black]{hyperref}
\usepackage{breakurl}  
\usepackage[mathscr]{euscript}
\usepackage{notes2bib}

\newcommand{\I}{\mathrm{i}}

\begin{document}

\title{Gouy's Phase Anomaly in Electron Waves Produced by Strong-Field Ionization}

\author{Simon Brennecke}
\affiliation{%
	Institut f\"ur Theoretische Physik, Leibniz Universit\"at Hannover, Appelstra{\ss}e 2, 30167 Hannover, Germany
}

\author{Nicolas Eicke}
\affiliation{%
	Institut f\"ur Theoretische Physik, Leibniz Universit\"at Hannover, Appelstra{\ss}e 2, 30167 Hannover, Germany
}

\author{Manfred Lein}%
\affiliation{%
Institut f\"ur Theoretische Physik, Leibniz Universit\"at Hannover, Appelstra{\ss}e 2, 30167 Hannover, Germany
}%

\date{\today}

\begin{abstract}
Ionization of atoms by linearly polarized strong laser fields produces cylindrically symmetric photoelectron momentum distributions that exhibit modulations due to the interference of outgoing electron trajectories. For a faithful modeling, it is essential to include previously overlooked phase jumps occurring when trajectories pass through focal points. Such phase jumps are known as Gouy's phase anomaly in optics or as Maslov phases in semiclassical theory. Most importantly, because of Coulomb focusing in three dimensions, one out of two trajectories in photoelectron holography goes through a focal point as it crosses the symmetry axis in momentum space. In addition, there exist observable Maslov phases already in two dimensions. Clustering algorithms enable us to implement a semiclassical model with the correct preexponential factor that affects both the weight and the phase of each trajectory. We also derive a simple rule to relate two-dimensional and three-dimensional models. It explains the shifted interference fringes and weaker high-energy yield in three dimensions. The results are in excellent agreement with solutions of the time-dependent Schr\"odinger equation.
\end{abstract}


\maketitle
The $\pi$ phase shift of an electromagnetic wave as it passes through a focus is an astonishing effect. Even though it was already observed by Gouy \cite{Gouy1890} more than 100 years ago, recent advances in laser technology have shone new light on Gouy's phase \cite{balcou1997,ruffin1999,lindner2004,andresen2013}. Analogous phenomena have also been found in other types of waves such as acoustic waves \cite{holme2003}, standing microwaves \cite{carpenter1959} and phonon-polariton wave packets in Raman scattering \cite{feurer2002}. In the past years, various experiments have been proposed and conducted for measuring the Gouy phase in matter waves \cite{dapaz2011,guzzinati2013,petersen2013}. 
In the present work, we demonstrate that Gouy's phase anomaly appears in electron wave packets produced by strong-field ionization leading to a significant imprint on interference structures in photoelectron momentum distributions (PMDs).

Strong-field ionization may be viewed as a two-step process consisting of: (i) release of an electron from the target; (ii) acceleration of the electron by the electromagnetic field in presence of the potential of the parent ion \cite{gallagher1988,corkum1989}. Depending on the system geometry, different parts of the emitted wave packets are mapped to the same final momentum, creating interference structures \cite{gopal2009,huismans2011,korneev2012}. As the positions of the interference fringes are determined by the phase difference between the wave packets, the emerging PMDs may be viewed as ``phasometer''. For linearly polarized laser pulses, the PMDs are dominated by photoelectron holography \cite{huismans2011,huismans2012}. 
Since these holograms \cite{gabor1992} may be qualitatively explained by the interference of a non-scattered ``reference'' wave packet and a scattered ``signal'' wave packet \cite{bian2011,bian2012}, they have been applied for ultrafast imaging \cite{meckel2014,haertelt2016,walt2017}.


While PMDs obtained by numerical solution of the time-dependent Schr\"odinger equation (TDSE) in full dimensionality (3D) agree well with
experimental data \cite{huismans2011,huismans2012}, PMDs obtained in reduced dimensionality (2D) are unable to reproduce quantitatively the fringe positions and emission strength of the experiment  \cite{meckel2014}. Indeed, by direct comparison we find that the positions of the holographic fringes in the numerical solution
of the TDSE in 3D differ from those in 2D. Due to the rotational symmetry in the 3D system (in contrast to 2D), the scattered wave packet is focused when it crosses the polarization axis (Coulomb focusing) \cite{brabec1996}. In analogy to Gouy's phase anomaly in optical waves, we show that the scattered wave packet acquires a $\pi/2$ phase shift in 3D compared to the 2D situation.


In order to provide an illustrative picture in terms of classical trajectories, several attempts were undertaken to describe PMDs in presence of long-range ionic potentials semiclassically. The resulting models are able to explain qualitatively the various types of interference structures for linearly polarized pulses \cite{popruzhenko2008,popruzhenko2008a,yan2010,li2014,shvetsov2016,lai2015,arbo2006}. However, since these models assume almost always that a slice through the 3D dynamics is equal to the corresponding 2D dynamics, they are unable to reproduce the positions of the holographic fringes in 3D and the relative weights of various kinds of trajectories. Putting aside the question of dimensionality, we note that approaches using the semiclassical phase based on Feynman's path integral \cite{lai2015,shvetsov2016} are promising since they lead to the correct number of holographic fingers and blades in the fan-like interference structure \cite{arbo2006,gopal2009,arbo2006a}. In a glory rescattering model the accurate description of the Coulomb-focused electrons in 3D was achieved \cite{xia2018}, but the interference between different types of trajectories was omitted. 

In this Letter, we show how to calculate semiclassical PMDs with quantitatively correct interference structures. Central ingredient is the preexponential factor of the semiclassical propagator \cite{levit1976,levit1977} that has been given little attention previously. Its modulus influences strongly the weights of the trajectories. Its phase, known as Maslov phase, affects the interference structures. We give a recipe how to analyze the focal-point structure of the system and calculate the corresponding Maslov phase. Using the rotational symmetry in linearly polarized fields, we formulate a simple rule how to evaluate the modulus and the phase in 3D compared to 2D. This additional phase is identified as a case of Gouy's phase anomaly.
For the analysis of the semiclassical trajectories, we propose a method that combines the shooting method \cite{huismans2011} with a clustering algorithm to determine all possible initial momenta that are classically mapped to a given final momentum (inversion problem). This enables us to go beyond~\cite{popruzhenko2008,popruzhenko2008a,yan2010,li2014,shvetsov2016} and to include the correct trajectory weights efficiently.



We calculate PMDs by numerical solution of the TDSE in single-active electron approximation using the split-operator method \cite{brennecke2019}. In 3D calculations, we choose an effective potential for helium \cite{tong2005} converted into a pseudopotential for the 1s state \cite{troullier1991}. In 2D calculations, this potential is further softened by replacing $r\rightarrow \sqrt{r^2+0.34}$, providing a 2D helium model with correct ionization potential $I_p\approx 0.9$~a.u. To obtain clean interference structures, we study the ionization dynamics for 2-cycle laser pulses polarized linearly along the $x$-axis as in \cite{brennecke2019}. Note, however, that the effects discussed below are also present for longer femtosecond pulses. 
\begin{figure}[t]
	\begin{center}
		\includegraphics[width=1.0\columnwidth,angle=0]{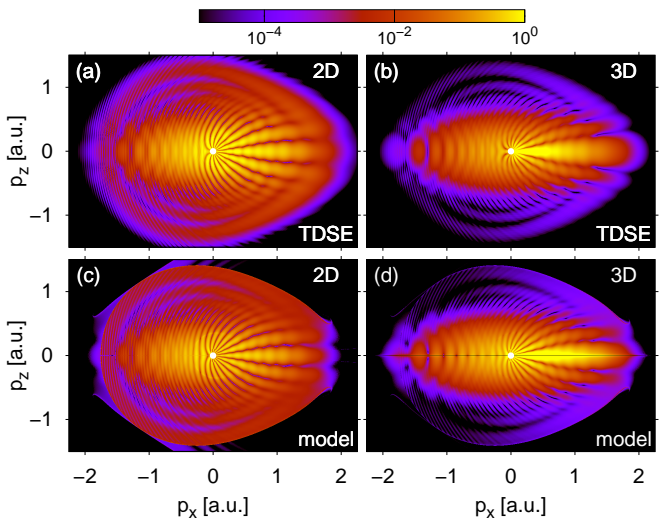}
		\vspace{-1.00cm}
	\end{center}
	\caption{Photoelectron momentum distribution in 2D [(a),(c)] and a slice along $p_y=0$ in 3D [(b),(d)] for helium ionized by a 2-cycle laser pulse with $4\times 10^{14}$ W/cm$^2$ intensity and 800~nm wavelength. The distributions are calculated by numerical solution of the TDSE in (a),(b) or with the semiclassical model in (c),(d).}  
	\label{fig1}
\end{figure}

The PMDs from solution of the TDSE are shown in Figs. \ref{fig1}(a),(b). Due to the rotational symmetry in 3D, it is sufficient to show a 2D slice through the 3D PMD at $p_y=0$. For $p_x<0$, ``direct'' non-scattered electrons  concentrated along the polarization axis dominate the distributions at momenta below the classical cut-off ($|p_x|\approx 1.63$~a.u.), whereas for $p_x>0$ holographic fringes roughly parallel to the polarization axis ($x$-axis) are visible. Both regions are overlaid with intracycle interferences roughly parallel to the $z$-axis \cite{gopal2009}. The weaker contribution of high-energy electrons, which extends over a large range of momenta $p_z$ (mostly $|p_z|>0.5$~a.u.), results from hard rescattering \cite{paulus1994,paulus1994a} and exhibits interference rings \cite{paulus2000,kopold2000}. Even though the 2D and 3D PMDs show the same overall structures, there are obvious differences in the positions of the interference fringes and in the relative strengths of different regions. 


Our semiclassical model is based on a combination of strong-field approximation (SFA) in saddle-point approximation (SPA) and semiclassical approximation to the time-dependent propagator formulated as path integral (see Ref. \cite{grossmann2008,kleinert2009} for a textbook treatment) similar to \cite{lai2015}. We neglect the ionic potential $V$ during the release of an initially bound electron so that this step may be described within SFA. Here, the initial canonical electron momentum $\textbf{p}'$ is conserved during the potential-free motion and would be equal to the final momentum, if the potential was also neglected in the second step. In order to reduce the workload we introduce connections between initial degrees of freedom such as in \cite{popruzhenko2008,popruzhenko2008a,yan2010}: For each initial momentum $\textbf{p}'$
the possible release times $t_0$ are given by the real part of the complex-valued SFA saddle-point times $t' = t_0 +\I t_t$. For its acceleration, the electron starts at the corresponding ``tunnel exit'' position given by $\textbf{r}_0=\mathrm{Re}(\int_{t'}^{t_0} \mathrm{d}\tau \textbf{A}(\tau))$ with an initial velocity $\dot{\textbf{r}}_0=\textbf{p}'+\textbf{A}(t_0)$. The acceleration of the electron from release time $t_0$ to final time $t_f\rightarrow \infty$ is modeled by a semiclassical approximation to the mixed position-momentum-space propagator $\langle \textbf{p}| U(t_f,t_0)| \textbf{r}_0\rangle$ such that the electric field and the Coulomb potential are included in a non-perturbative manner. This involves trajectories satisfying classical equation of motions (EOM) governed by the Hamiltonian
$H(\textbf{r},\textbf{p},t)=\frac{1}{2}\left(\textbf{p}+\textbf{A}(t)\right)^2+V(\textbf{r})$
with the electric field defined by $\textbf{E}(t)=-\partial_t \textbf{A}(t)$.

The PMD is proportional to the modulus squared of 
the transition amplitude which can be approximated in terms of trajectories as (van Vleck approximation)
\begin{equation}
M_{\textbf{p}}(t_f)\propto  \sum_{s} \, \frac{D\,C_{\mathrm{Coul}}}{\sqrt{|J|}}\,
e^{\I (S^0_{\downarrow}+S_{\rightarrow}-\nu\pi/2)},
\label{eq:finalamplitude}
\end{equation} 
where $s$ labels all possible initial momenta $\textbf{p}'$ that are classically mapped to the given final momentum $\textbf{p}$. The matrix element $D$ is the prefactor from SFA in SPA and $C^2_{\mathrm{Coul}}=\left(2\kappa^2/|\textbf{E}(t_0)|\right)^{2/\kappa}$ with $\kappa=\sqrt{2I_p}$ is the Coulomb correction of the ionization rate \cite{perelomov1967,bisgaard2004,popruzhenko2008}. The action associated to each trajectory may be split into a complex-valued part corresponding to the ionization step
\begin{equation}
S^0_{\downarrow}=I_pt'-\int_{t'}^{t_0}\mathrm{d}t\, \frac{(\textbf{p}'+\textbf{A}(t))^2}{2}
\end{equation} 
and a part corresponding to the acceleration step \cite{shvetsov2016}
\begin{equation}
S_{\rightarrow}=-\int_{t_0}^{\infty}\mathrm{d}t\,\biggl[\frac{1}{2}\dot{\textbf{r}}^2(t)+V(\textbf{r}(t))-\textbf{r}(t)\cdot \nabla V(\textbf{r}(t))\biggr].
\label{eq:semiphase}
\end{equation}
The reduced propagator is completely determined by the Jacobian fields $\partial \textbf{p}(t)/\partial \textbf{p}'$, $\partial \textbf{r}(t)/\partial \textbf{p}'$, also known as stability matrices \cite{levit1976,levit1977}. These are calculated 
by integration of the Jacobi initial value problem as described in \cite{levit1976,kay1994}. The Jacobian $J$ governs the change of the classical phase-space density, 
\begin{equation}
J=J(t_f)=\mathrm{det}\left(\frac{\partial \textbf{p}(t_f)}{\partial \textbf{p}'}\right).
\label{eq:jaco}
\end{equation}
The Maslov index $\nu$ may be viewed as a time-dependent function $\nu(t)$ with $\nu(t_0)=0$ that can only change at times $T$ with vanishing Jacobian $J(T)=0$. At these points $T$ there are $m$ linearly independent zero modes $\textbf{d}^{(i)}$ of $\partial\textbf{p}(T)/\partial \textbf{p}'$. Since infinitesimal changes of the initial
momenta $\textbf{p}'\rightarrow \textbf{p}'+\epsilon\, \textbf{d}^{(i)}$ in these directions $\textbf{d}^{(i)}$ do not change the momentum $\textbf{p}(T)$ in first order of $\epsilon$, the points $T$ are called focal points. The corresponding infinitesimal changes $\delta \textbf{r}^{(i)}$ of the position are given by 
\begin{equation}
\delta \textbf{r}^{(i)}=\epsilon \sum_j \frac{\partial \textbf{r}(T)}{\partial p'_j} d^{(i)}_{j}.
\end{equation} 
The jump of the Maslov index across focal points is
\begin{equation}
\Delta \nu(T)=m-1+\mathrm{sgn}\, \mathrm{det}(g),
\label{eq:maslovindex}
\end{equation}
where the matrix elements of the $m\times m$ matrix $g$ are given by $g_{i,j}=\delta \textbf{r}^{(i)}\cdot \mathrm{Hesse}_{\textbf{r},\textbf{r}} (H)\delta \textbf{r}^{(j)}$ with the Hessian matrix $\mathrm{Hesse}_{\textbf{r},\textbf{r}} (H)$ of the Hamiltonian $H$ with respect to the positions $\textbf{r}$ evaluated along the classical trajectory~\cite{levit1977}. In position representation, $\Delta \nu$ is equal to the multiplicity $m$ of the focal point, which is known as Morse theorem \cite{morse1973}. In mixed representation used here, this is not generally true. However, if the Hesse matrix $\mathrm{Hesse}_{\textbf{r},\textbf{r}} (H)$ is positive definite, Eq.~\eqref{eq:maslovindex} reduces to the Morse theorem.

In order to solve the inversion problem, a Monte-Carlo algorithm is used to sample a large number of possible initial momenta $\textbf{p}'$. The corresponding final momenta $\textbf{p}$ are binned in cells in momentum space.
For a large number $n_\textbf{k}$ of trajectories per bin, the solutions in a single bin form accumulations in the space of initial momenta which may be identified by means of a clustering algorithm \footnote{We use a DBSCAN (density-based spatial clustering of applications with noise) algorithm implemented in the scikit-learn package for Python.}. In the limit of vanishing bin size $\Delta k_i\rightarrow0$, the area covered by the clusters goes to zero and the locations of the clusters are the solutions of the inversion problem for a final momentum $\textbf{k}$. This method enables us to determine all possible solutions in a systematic manner, which is a difficult problem in other approaches to solving the inversion problem \cite{lai2015}. The area covered by the clusters is proportional to $1/|J|$, so that the shooting methods \cite{huismans2011,shvetsov2016,li2014} imply a wrong weighting of trajectories. Solving the inversion problem allows us to determine the correct weight  $1/\sqrt{|J|}$ in Eq.~\eqref{eq:finalamplitude} by explicit calculation for each type of trajectories.
\begin{figure}[t]
	\begin{center}
		\includegraphics[width=1.0\columnwidth,angle=0]{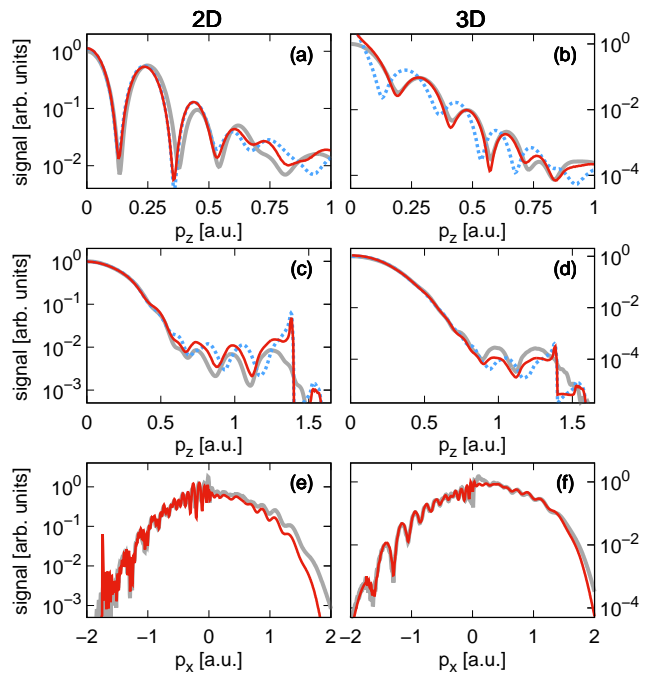}
		\vspace{-1.00cm}
	\end{center}
	\caption{(a)-(d) Slices through PMDs of Fig. \ref{fig1} at fixed $p_x=E_0/(2\omega)\approx0.81$~a.u. in (a),(b) and at fixed $p_x=-0.5$~a.u. in (c),(d). (e),(f) Longitudinal distribution obtained by integration over $p_\perp\le0.2$~a.u. Calculations in 2D and 3D are displayed in the left and right panels, respectively. TDSE result (gray thick line), semiclassical model including Maslov's phase (red thin line) or neglecting Maslov's phase (blue dotted line).}  
	\label{fig2}
\end{figure}


The PMDs calculated with the semiclassical model for a bare Coulomb potential $-1/r$ are shown in Figs.~\ref{fig1}(c),(d). Our model reproduces perfectly the 2D and 3D TDSE distributions as the Jacobian fields depend on the dimension. Using the symmetry, a relation between the Jacobian $J_{3D}$ of the 3D system and the Jacobian $J_{2D}$ of the corresponding 2D system is obtained,
\begin{equation}
|J_{3D}|=\frac{p_\perp}{{p}_\perp'} |J_{2D}|\quad\text{with}\quad p_\perp=\sqrt{p_y^2+p_z^2}.
\label{eq:jacobian}
\end{equation} 
This 2D-to-3D correction weight has already been used in~\cite{shvetsov2016}.  
For ``direct'' non-scattered trajectories with $p_\perp\approx p_\perp'$ the Jacobians in both dimensionalities are nearly the same. In contrast, rescattering trajectories with large scattering angles correspond to small initial perpendicular velocity $p_\perp'\ll p_\perp$ such that their Jacobian is enhanced and hence their weighting reduced in 3D compared to 2D. Physically, this reflects the increased recollision probability in 2D compared to 3D. Close to the polarization axis, where $p_\perp\approx0$, the distributions in 3D are greatly enhanced due to Coulomb-focusing with $p_\perp'\gg p_\perp$ \cite{brabec1996,xia2018}. Hence, the difference in the Jacobian $J$ explains the higher emission strength of electrons with large $p_\perp$ in 2D compared to 3D, see also the different scales of the 1D slices shown in Figs. \ref{fig2}(a)-(d), and the different modulation depth of the holographic fringes. 
The longitudinal momentum distributions, shown in Figs. \ref{fig2}(e),(f), are also in good agreement for both levels of theory except close to the classical cut-off where neglected laser-driven recollisions under the Coulomb barrier are important~\cite{keil2016}.
%
\begin{figure}[t]
	\begin{center}
		\includegraphics[width=1.0\columnwidth,angle=0]{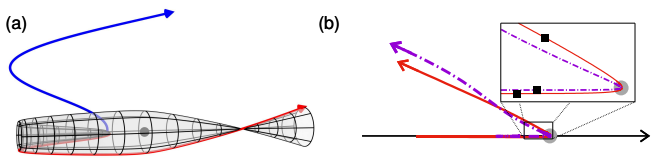}
		\vspace{-1.00cm}
	\end{center}
	\caption{(a) Photoelectron holography: interference between non-scattered trajectory (blue line) and forward scattered trajectory (red line). Due to the rotational symmetry in 3D, a one dimensional manifold of trajectories with the same initial perpendicular momentum $p_\perp'$ (indicated as surface) is focused on the polarization axis resulting in a phase jump of $\pi/2$. (b) High-order above-threshold ionization: interference between long (red solid line) and short (violet dashed-dotted line) rescattering trajectories. The short trajectory goes through a single focal point (black point) such that the Maslov index $\nu=+1$ whereas the long trajectory passes two focal points (black points) resulting in Maslov index $\nu=0$.}  
	\label{fig3}
\end{figure}

\begin{figure}[t]
	\begin{center}
		\includegraphics[width=0.9\columnwidth,angle=0]{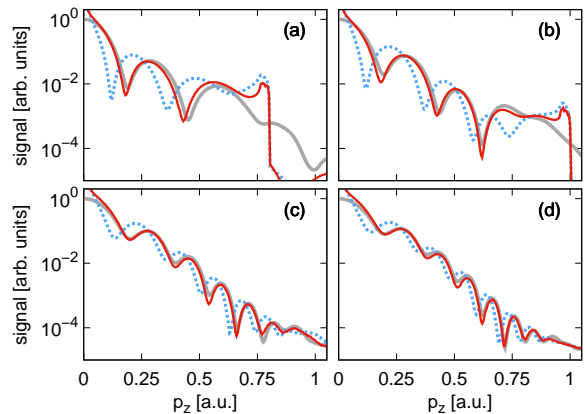}
		\vspace{-0.60cm}
	\end{center}
	\caption{1D Slice through 3D PMDs at fixed $p_x=E_0/(2\omega)$ as in Fig. \ref{fig2}(b) but for various wavelengths: (a) $\lambda=400$~nm, (b) $\lambda=600$~nm, (c) $\lambda=1000$~nm and (d) $\lambda=1200$~nm. TDSE result (gray thick line), semiclassical model including Maslov's phase (red thin line) or neglecting Maslov's phase (blue dotted line).}
	
	\label{fig4}
\end{figure}

The focal-point structure is also modified in 3D compared to 2D because zeros of $p_\perp(T)$ along the trajectory lead to vanishing Jacobian $J_{3D}$ of Eq.~\eqref{eq:jacobian}. The rotational symmetry dictates that in such points, all trajectories with the same magnitude of their initial perpendicular velocity and the same release time $t_0$ are mapped to $p_\perp(T)=0$ and hence they are focused in momentum space. In an alternative view, trajectories crossing the polarization axis $r_\perp=0$ are focused in position space so that this quantum-mechanical system is similar to an optical cylindrical lens \cite{mcgowan2000}. The situation is schematically illustrated in Fig.~\ref{fig3}(a). The corresponding Gouy phase anomaly is mathematically represented by the different Maslov indices in 2D and 3D: for potentials $V(x,r_\perp)$ with a minimum at $r_\perp=0$ and arbitrary trajectories it follows from  Eq. \eqref{eq:maslovindex} that the indices are related by
\begin{equation}
\nu_{3D}=\nu_{2D}+\delta \nu,
\label{eq:maslov3D}
\end{equation} 
where $\delta \nu$ is the number of zero crossings of $p_\perp(T)$ along the trajectory. The holographic pattern can be understood as interference of a non-scattered ``reference'' wave and a scattered ``signal'' wave \cite{huismans2011,bian2011}. For final momenta with $p_y=0$ and $p_z>0$, the ``reference'' trajectory has a perpendicular initial velocity $p_z'>0$ and it is only weakly perturbed by the potential. Hence its Maslov phase is 0. In contrast, the ``signal'' trajectory starts with initial velocity $p_z'<0$ and it is strongly deflected by the potential to its final momentum $p_z>0$. The explicit calculation shows that the Maslov index in 2D is still 0. However, since in 3D an additional axial focal point is crossed, the Maslov phase jumps to $\pi/2$, see Fig.~\ref{fig3}(a). This phase jump affects the positions of minima and maxima, see the 1D slices through the holographic ``fingers'' shown in Figs. \ref{fig2}(a),(b). The inclusion of Maslov's phase and hence the consideration of Gouy's phase anomaly enables us to explain the exact fringe positions in the lateral momentum distribution in 3D for various different wavelengths as shown in Fig.~\ref{fig4}.

Effectively, a 2D semiclassical model can be turned into a 3D model by multiplying the complex amplitude of each trajectory with the factor $\sqrt{p'_\perp/p_\perp}\exp(-\I\,\delta \nu\, \frac{\pi}{2})$.
In addition to this correction, there are observable non-vanishing Maslov phases already in 2D. As these phases result from passing through other types of caustics, they have to be calculated numerically using Eq. \eqref{eq:maslovindex}. One example concerns the nearly circular ring structures at high energies arising from interference of short and long rescattering trajectories \cite{paulus2000,kopold2000}. Even though the long trajectory has $\nu_{2D}=0$, it passes two focal points in 2D: at one point the Maslov index is increased by $+1$ and at the other it is decreased by $-1$. In contrast, the short trajectory only crosses a single focal point in 2D resulting in a Maslov index $\nu_{2D}=+1$, compare Fig. \ref{fig3}(b) for a schematic illustration. The resulting phase difference $\pi/2$ is essential to reproduce the correct positions of the circular ring structures which can be seen in the 1D slices of Figs. \ref{fig2}(c),(d). In quantum-orbit models based on SFA \cite{paulus2000,kopold2000}, this additional phase is automatically included by the Hessian matrix element in the prefactor of the SPA, so that the positions of the interference rings are correct (not shown)~\cite{brennecke2018a}.  

To conclude, we have shown that Gouy's phase anomaly in electron waves plays a central role in the formation of interference structures in PMDs from strong-field ionization. For linearly polarized fields in 3D, Coulomb focusing leads to the appearance of focal points that are absent in 2D, so that the holographic interference fringes are shifted and the spectral weight is modified. Our quantitative interpretation of the PMDs from atoms by means of a powerful semiclassical model forms the foundation for the future analysis of more complex system with e.g. non-trivially polarized fields \cite{mancuso2016,eckart2018,li2018} or more complex targets such as molecules \cite{Doblhoff2013,haertelt2016,liu2016,walt2017,morishita2019}. We expect that the changed focal-point structure influences decisively the resulting PMDs.
For the full agreement of our model with \textit{ab-initio} TDSE results, the inclusion of preexponential factors is crucial. We have accomplished this task by using an efficient clustering method for solving the inversion problem and have thus shown an alternative approach to resolve one of the problems in the applicability of the van Vleck propagator. This idea can be transferred to other semiclassically-treated problems, e.g. in the context of molecular spectroscopy \cite{kay1994,kay2005,conte2019} or interacting bosons \cite{ray2016}, with the advantage of a clearer view on physical processes compared to previously used initial-value representations.

\begin{acknowledgments}
We thank Nikolay Shvetsov-Shilovski for valuable discussions. 
This work has been supported by the Deutsche Forschungsgemeinschaft through the Programme \textit{Quantum Dynamics in Tailored Intense Fields} (QUTIF).
\end{acknowledgments}


%

\end{document}